\documentclass[twocolumn,showpacs,preprintnumbers,amsmath,amssymb]{revtex4}
\usepackage{graphicx}
\begin{document}
\title{ $d+id'$-wave Superconducting States in Graphene}
%EC:  hexagonal is really triangular
\author{ Yongjin Jiang$^{1,2}$, Dao-Xin Yao$^2$, E. W. Carlson$^{2}$, Han-Dong Chen$^3$ and JiangPing Hu$^2$}
\affiliation{$^1$Department of Physics, ZheJiang Normal University,
Jinhua, Zhejiang, 321004,P.R.China} \affiliation{$^2$Department of
Physics, Purdue University, West Lafayette, IN, USA}
\affiliation{$^3$Department of Physics, University of Illinois at
Urbana-Champaign, Urbana, IL 61801, USA}
\begin{abstract}
   We show that effective superconducting orders generally emerge at low energy in the superconducting state of graphene with conventionally defined pairing symmetry .
   We study such a particular interesting example, the $d_{x^2-y^2}+id'_{xy}$ spin singlet pairing superconducting state  in
   graphene, which can be generated by electronic correlation as well as induced
   through a proximity effect with a d-wave superconductor. We find that effectively the d-wave state is
a state with mixed s-wave and exotic $p+ip$-wave pairing orders at
low energy. This remarkable property leads to distinctive
superconducting gap functions  and novel behavior of the Andreev
conductance spectra.
\end{abstract}
 \maketitle \maketitle

  Graphene is a single layer of hexagonally
  %EC packed
  coordinated carbon atoms which has recently been isolated\cite{first}.
  Due to its special lattice structure, the low energy part of its energy spectrum is
  characterized
  by particle-hole symmetric linear dispersions
  around  the corners of the hexagonal Brilliouin zone (BZ). This
  %EC property
  band structure is responsible for
  many new properties of this `relativistic' condensed
  matter system,
  such as an abnormal quantum Hall effect\cite{expnature,theo1,theo2},
  minimum conductance\cite{theo2,miniconductancetheo}, and
  possibly even an
experimental realization of the Klein paradox\cite{klein}.

   Recently,
  a novel concept called specular
  Andreev reflection was proposed for a normal/superconducting(N/S) graphene
 %EC structure
 interface in the context of a conventional
  s-wave pairing superconducting state\cite{Beenakker1}.
  Later, an unusual oscillation of the quantum conductance through
  an N/I/S junction was predicted\cite{swaveosi,dwaveosi}.
  The possible superconducting pairing orders have also been studied.
  In
Ref.~\onlinecite{d1}, by including strong electronic correlations,
the mean field search   shows that $d_{x^2-y^2}+id_{xy}'$-wave
pairing symmetry is favored, similar to the superconducting state in
the triangular lattice which is believed to be of
$d_{x^2-y^2}+id_{xy}'$
%EC pairing type
symmetry\cite{tri1,tri2}. In Ref.~\onlinecite{p+ip},
  an exotic $p+ip$-wave
  superconductor with spin singlet bond pairing
 was suggested at the mean field level and
 possible phonon- or plasma-mediated mechanisms were discussed.
  On the other hand, experimentally,
superconducting states in graphene have been realized by proximity
effect\cite{andreevexp,natureexp,expevidence} through contact with
superconducting electrodes.

\begin{figure}[tbh]
\includegraphics[width=9cm,height=8cm]{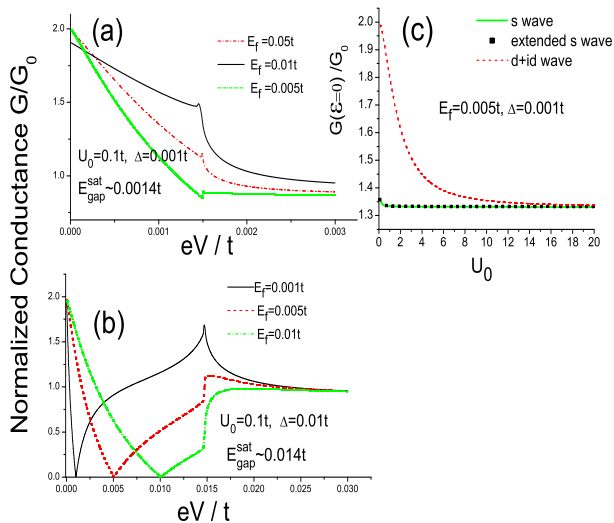}
% Here is how to import EPS art
\caption{The normalized quantum conductance of a S/N graphene
junction is shown in (a) with $E_f>E_{\rm gap}^{sat}$ and (b) with
$E_f<E_{\rm gap}^{sat}$ for heavily doped
superconductor(U=0.1t,i.e., ~ 300 meV). Also shown in (c) the
normalized conductance for zero bias voltage for three kinds of
pairing order parameters, i.e., conventional s-wave, extended
s-wave(bond pairing) and $d+id'$ wave(bond pairing)}
 \label{fig:2}
\end{figure}

The peculiar  physics in graphene is  the unusual linear
and isotropic dispersion of the low energy excitations
around the Dirac points.  In this Letter, we show that because of
the existence of the Dirac points,
% EC rearranged this sentence
conventional pairing order parameters can lead
to the emergence of exotic pairing states
in the low energy effective description.
%effective superconducting
%orders emerge at low energy  in the superconducting state with
%conventional pairing order parameters.
The p+ip superconducting order of Ref.~\onlinecite{p+ip} is
precisely such an example of an effective, low energy
superconducting order, arising in that case from a more
conventional extended s-wave pairing. Here, we study a
particularly interesting superconducting state in graphene, the
$d_{x^2-y^2}+id_{xy}$ spin singlet pairing superconducting state,
which  can be generated by electronic correlation\cite{d1} as well
as induced through a proximity effect with a d-wave superconductor
on top of
%EC
or underneath the graphene layer. We find that the d-wave state is
effectively a mixed s-wave and exotic $p+ip$-wave pairing state at
low energy.
%EC
The mixture of both s-wave and $p+ip$ wave leads to unique
properties of the excitation spectrum and Andreev conductance
spectra. The excitation spectrum is gapless at half-filling  and
is  gapped away from half filling.
%EC
The gap is equal to the chemical potential near half filling, and
it saturates as the chemical potiential is moved above the energy
scale set by pairing strength.
%The size of the gap is equal to the chemical potential near half filling
%below the energy scale set by the pairing and remain roughly
%unchanged for larger chemical potential.
The
%EC
normalized
Andreev conductance in
the limit of zero bias voltage is a smooth function of the
chemical potential, which starts from $2$ at half filling and  drops
smoothly to $4/3$  at large doping, unlike that in the s-wave
pairing states where it almost remains  at a constant value, $4/3$
(see Fig.~\ref{fig:2}(c)).
%EC
This is a signature of $d_{x^2-y^2}+id'_{xy}$ pairing in graphene.
%The result provides signature to experimentally detect possible
%$d_{x^2-y^2}+id_{xy}$ superconducting state in graphene.
%***% Fig.~\ref{fig:2}(c) is mislabeled -- switch d+id and extended s in the legend ***

\begin{figure}[tbh]
\includegraphics[width=6cm,height=4cm]{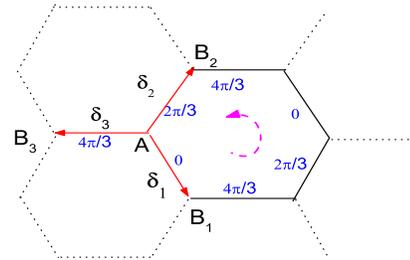}
% Here is how to import EPS art
\caption{(Color online.) Phase (blue number) of singlet bond
pairing function on the graphene lattice which
%EC keeps translation symmetry and is
preserves the translational and rotational symmetry of the
honeycomb lattice and is $d+id'$ type under point group $D_{6}$.
The red vectors $\vec \delta_a$(a=1,2,3) denote nearest-neighbor
inter-sublattice connections.} \label{fig:0}
\end{figure}

{\it General pairing symmetry in graphene:}  Although the crystal
point group of graphene is $D_{6h}$,  the pairing symmetry of the
superconducting orders in a two dimensional graphene  sheet is
governed by  $D_6$, which includes four one-dimensional irreducible
representations, $A_{1,2}$ and $B_{1,2}$, and two two-dimensional
irreducible representations, $E_{1,2}$.  Among these
representations, the $A_1, E_1$ and $E_2$  representations describe  s-wave,
p-wave and  d-wave pairing symmetries, respectively.  Therefore, the
spin singlet s-wave  and d-wave pairing are described by the $A_1$
and $E_2$ irreducible representations.
We can understand the pairing symmetry further by
considering the exchange symmetry between the A and B sublattices.
The $D_6$ group is a direct product of its two subgroups $C_{3v}$
and $Z_2$, i.e. $D_6=C_{3v}\bigotimes Z_2$, where $Z_2$  describes
the exchange operations between the A and B sublattices.  The $A_1$
and $E_2$ representations of $D_6$ are symmetric under  exchange
of the A and B sublattices, while the $E_1$
representation is antisymmetric.
%EC
%Note that parity is a good quantum number due to inversion symmetry
%(equivalent to the $Z_2$ sublattice symmetry),
%and singlet and triplet pairing will not mix.

{\it Emergent pairing symmetry at low energy:} At low energy, the
effective physics in graphene can be described by a relativistic
dispersion near the wave vectors $ \vec
K_{\pm}=(0,\pm\frac{4\pi}{3\sqrt{3}})$( hereafter subscript
$'\pm'$ always denotes the valley index). In the superconducting
state of graphene, we also have to consider the  superconducting
orders near these vectors at low energy,  which leads to the
effective superconducting orders. In particular, when the pairing
is between two sublattices, the effective superconducting orders
can have new pairing symmetry around the Dirac cones.
%EC For a simple consideration, let's
To see this, consider a translationally invariant superconducting order
defined on the links of the nearest neighbor sites between the A
and B sublattices.
In real space, this pairing order
is described by three independent values $(\Delta_{\vec
\delta_1},\Delta_{\vec \delta_2},\Delta_{\vec \delta_3})$ as shown
in Fig.\ref{fig:0}.  In momentum space, the superconducting order
is given by
\begin{eqnarray}
\Delta(\vec k)=\sum_a \Delta_{\vec \delta_a} e^{i\vec k\cdot \vec
\delta_a} \label{eq:det}
\end{eqnarray}
At low energy, near the Dirac cones, the effective
superconducting order is given by $\Delta_{\pm}(\vec q)=\Delta(\vec
K_{\pm}+\vec q)$. Given a small $\vec q$,  we have
\begin{eqnarray}
\label{effective}
 \Delta_{\pm}(\vec q) =\Delta(\vec K_{\pm})+i \vec
q \cdot (\sum_a\vec \delta_a\Delta_{\vec \delta_a} e^{\pm i\vec
K_{\pm}\cdot \vec \delta_a}).
\end{eqnarray}
Let us consider two specific cases. The first case is extended
s-wave pairing. In this case, $\Delta_{\vec \delta_a}=\Delta,
a=1,2,3$.  The first term in the right side of
Eq.(\ref{effective}) vanishes and it is easy to show that
$\Delta^s_{\pm}(\vec q)=-\frac{3}{2}\Delta(\pm q_y+iq_x)$, which
becomes a $p$-wave like pairing order. Therefore, the extended
s-wave pairing order in graphene at low energy is described by two
$p+ip$ pairing orders that are connected with each other by time
reversal symmetry. This case has been studied in
Ref.~\onlinecite{p+ip}. The second case is  $d_{x^2-y^2}+id'_{xy}$
wave pairing on which this paper is focused.  In this case,
$\Delta_{\vec \delta_i}=\Delta e^{2i a \pi/3}, a=1,2,3.$ The
effective superconducting orders for small $\vec q$ in
Eq.(\ref{effective}) become
\begin{eqnarray}
\Delta^d_{+}(\vec q)= 3\Delta e^{i\frac{4\pi}{3}} \nonumber\\
\Delta^d_{-}(\vec q)=\frac{3}{2}e^{i\frac{\pi}{3}}\Delta(iq_x+q_y)
\label{eq:3}
\end{eqnarray}
%EC
The first equation, $\Delta^d_{+}(\vec q)$, corresponds to
$s$-wave pairing, and the second, $\Delta^d_{-}(\vec q)$, to
$p+ip$-wave pairing.
%The first order $\Delta^d_{+}(\vec q)$ is a s-wave pairing and the
%second order $\Delta^d_{-}(\vec q)$ is a $p+ip'$ wave pairing.
Therefore, {\em at low energy, the $d_{x^2-y^2}+id'_{xy}$ wave
pairing state in graphene is a superconducting state with mixed
$s$ and $p+ip$ pairing orders.}

{\it Lattice model and the quasi-particle spectrum in mean
field:}   The graphene system is composed of two sublattices which
are labelled as A and B,  as shown in Fig.\ref{fig:0}. If
the superconducting pairing  is betwen two sublattices, the pairing
Hamiltonian can be written at the mean field level as follows,
\begin{eqnarray}
H=&-t&\sum_{i,a,\sigma} [A_{i\sigma}^{\dagger}B_{i+\vec \delta_a\sigma} + H.C.] \nonumber \\
&+& \sum_{i,a}[\Delta_{\vec \delta_a}(A_{i\uparrow}^{\dagger}B_{i+\vec \delta_a\downarrow}^{\dagger}-A_{i\downarrow}^{\dagger}B_{i+\vec \delta_a\uparrow}^{\dagger}) + H.C.]      \nonumber\\
&-& \mu\sum_{i,\sigma}(A_{i\sigma}^{\dagger}A_{i\sigma}+B_{i+\vec
\delta_1 \sigma}^{\dagger}B_{i+\vec \delta_1 \sigma}),
\label{eq:1}
\end{eqnarray}
where the index $i$ sums over sites on the A sublattice.
$A_{i\sigma}^{\dagger}$ and $B_{j\sigma}^{\dagger}$ are creation
operators for two sublattices and $\sigma=\uparrow,\downarrow$ are
spin indices. The first term describes free band where $t\sim
2.8eV$ is the nearest-neighbor hopping constant.  In the pairing term,
 $\Delta_{\vec \delta_a}$ is the spin singlet
bond pairing order parameter which
%EC keeps translation symmetry and
has $d+id'$ symmetry under the point group $D_6$, {\em i.e.},
$\Delta_{\vec \delta_a}=\Delta e^{2i a \pi/3}$, where $\Delta$ is
the pairing strength. The phase of the order parameter winds by $4
\pi$ around each hexagonal plaquette(shown in Fig.\ref{fig:0}).
This ansatz preserves the rotational
 and translational
 symmetry of the
 original lattice but breaks time reversal symmetry(TRS) manifestly.  The chemical potential, $\mu$,
can be tuned by the gate voltage.

In momentum space, we can rewrite the Hamiltonian in the form:
$H=\sum_{\vec{k}}\Psi_{\vec{k}}^{\dagger}\tilde{H}_{\vec
k}\Psi_{\vec{k}}+const$, where we defined the Nambu spinor
$\Psi_{\vec{k}}=(A_{\vec k\uparrow},B_{\vec k\uparrow},A_{-\vec
k\downarrow}^{\dagger},B_{-\vec k\downarrow}^{\dagger})$
 and the 4$\times$ 4 matrix $\tilde{H}_{\vec k}$ is:
\begin{equation}
\tilde{H}_{\vec k}= \left(
\begin{array}{cccc}
-\mu & f(\vec k)& 0 & \Delta(\vec k)\\
f(\vec k)^{*} & -\mu& \Delta(-\vec k)&0\\
0 &\Delta(-\vec k)^{*}&\mu&-f(-\vec k)^{*}\\
\Delta(\vec k)^{*}& 0&-f(-\vec k)&\mu\\
\end{array}\right)\
\label{eq:4}
\end{equation}
where the function $f(\vec{k})$ is defined by
$f(\vec{k})=-t\sum_{a}e^{i\vec{k}\cdot\vec \delta_a}$ and
$\Delta(\vec k)$ is defined by Eq.(\ref{eq:det}).

At low energy, we can linearize the mean field Hamiltonian
Eq.(\ref{eq:4}) near the two inequivalent BZ corners $\vec K_{\pm}$.
%EC It can  be obtained that
 Near $K_{\pm}$, $f(\vec K_{\pm}+\vec{k})$
 can be expanded as:
 \begin{equation}
f_{\pm}(\vec{k})=f(\vec K_{\pm}+\vec{k})=v(ik_x \pm k_y) \\
\label{eq:6}
 \end{equation}
where we introduced a valley dependent function $f_{\pm}$.
The velocity of the Dirac particles is $v=\frac{3t}{2}$. Note hereafter
$k_x,k_y$ always refer to the relative vectors measured from
$K_{\pm}$.  By substituting Eq.(\ref{eq:3}) and Eq.(\ref{eq:6}) into
 Eq.(\ref{eq:4}),  we obtain
\begin{equation}
\tilde{H}_{\pm}(\vec{k})= \left(
\begin{array}{cccc}
v(\pm k_y\sigma_x - k_x\sigma_y)-\mu & \tilde{\Delta}_{\pm}(\vec k)\\
\tilde{\Delta}_{\pm}^\dagger(\vec k) & \mu-v(\pm k_y\sigma_x - k_x\sigma_y)\\
\end{array}\right)\;
\label{eq:11}
\end{equation}
where $\sigma_{x,y}$ refer to the Pauli matrices. The linearized
pairing matrices $\tilde{\Delta}_{\pm}(\vec k)$ for the  $'\pm'$ valleys
take the form,
\begin{equation}
\tilde{\Delta}_{+}(\vec k)= \Delta\left(
\begin{array}{cccc}
0&3e^{i\frac{4\pi}{3}}\\
\frac{3}{2}(-ik_x-k_y)e^{i\frac{\pi}{3}} & 0\\
\end{array}\right)\
\end{equation}
and $\tilde{\Delta}_{-}(\vec k)=\tilde{\Delta}_{+}(-\vec k)^{T}$. It
can be easily checked that $\tilde{H}_{+}(\vec{k})$ and
$\tilde{H}_{-}(-\vec{k})$ transform into each other under spacial
inversion, $A_{\vec k}\rightleftharpoons B_{-\vec k}$. Here we note
that the $d+id'$ pairing ansatz break TRS but preserves inversion symmetry
so that  the valley degeneracy is unbroken.

\begin{figure}[tbh]
\includegraphics[width=7cm,height=5cm]{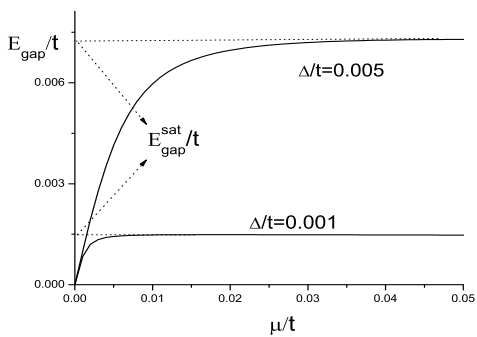}
% Here is how to import EPS art
\caption{The energy gap $E_{\rm gap}$ as a function of chemical
potential for pairing strength $\Delta$=0.001t and
$\Delta$=0.005t, i.e., $\sim$3mev and $\sim$15mev separately.
Notably, $E_{\rm gap}$ is linear at low doping region and
saturates to a constant $E_{\rm gap}^{sat}$ when $\mu\gg\Delta$.}
\label{fig:1}
\end{figure}

The elementary excitation spectrum can be obtained through
Bogoliubov diagonalization. Furthermore, we can find the energy gap
$E_{\rm gap}$
%EC corresponds to minimum excitation energy.
corresponding to the minimum excitation energy. It can be shown
rigorously that for $\mu\ll\Delta$, $E_{\rm gap}=\mu$. In
Fig.~\ref{fig:1} we plot the gap as a function of chemical
potential. $E_{\rm gap}$ is linear in the $\mu\ll\Delta$ (low
doping) region and saturates to a constant $E_{\rm gap}^{sat}$ for
$\mu\gg\Delta$. This unique dependence of the energy gap on the
chemical potential in the $d+id'$ superconducting state stems from
the mixture of the s-wave and the $p+ip$ wave components. Since
the $p+ip$ wave component dominates at low doping, the gap depends
linearly  on the chemical potential.  Since  the s-wave component
dominates in the high doping region, the gap saturates above the
s-wave superconducting order parameter strength.

{\it Andreev conductance through S/N junction:} In the following,
we show that the mixture of the s-wave and $p+ip$-wave in the
$d+id'$ superconducting state of graphene results in
%EC distinguished
a distinctive signature in the
Andreev conductance.  Consider a S/N
 graphene junction with the x$<$0 region being the graphene $d+id'$ superconductor
 and the x$>$0 region being the normal state of graphene.
 %EC  metal. -- it's a semimetal
  We assume that the  electrostatic potential on the S side is lower than
that on the N side
  by a value $U_0>0$, which can be fixed through the gate voltage or by doping. A large
$U_0$ implies a heavily doped superconductor.
  %The Fermi energy $E_F$ is the same on both sides and is a tunable parameter. With these
%parameters, the chemical potential $\mu=E_F$ on the N side and
%$\mu=E_F+U_0$ on the S side.
%***EC:  Odd way to say it.  Usually the chemical potential is defined to be the same for two materials in contact. ***
 Due to the
 spin and valley degeneracy, we can restrict the incident state
 from N side to be spin up and from valley $'$+$'$ and multiply the
 conductance by 4 at the end.

Under certain voltage bias V, we expect the incidence of a particle
excitation with energy $\varepsilon=eV$ from the x$>$0 side into the
junction at x=0. The general form of the incident wave function is
$\Psi^e_{i}=\Phi^e(-k_x,k_y) e^{i(-k_xx+k_yy)}$ where $k_x( k_y)$ is
the longitudinal(transverse) component of the wave vector. In the
scattering process, we assume energy and the transverse component of
the wave vector is conserved. The reflected states can be either an
electron state $\Psi^e_{r}$=$\Phi^e(k_x,k_y) e^{i(k_xx+k_yy)}$ or a
hole state $\Psi^h_{r}$=$\Phi^h(k'_x,k_y) e^{i(k'_xx+k_yy)}$ where
$k'_x$ is determined by $vk'$=$|\varepsilon-E_F|$, it is negative
for $\varepsilon<E_F$(retro-reflection) and positive for
$\varepsilon>E_F$(specular reflection)\cite{Beenakker1}. It can also
be imaginary if $vk_y>|\varepsilon-E_F|$ and the corresponding hole
state is an evanescent state near the boundary . $\Phi^e$ and
$\Phi^h$ are 4-component spinor eigenstates
 of Eq.(\ref{eq:11}) on N side(for which $\Delta$=0) corresponding to electron and
 hole excitations, respectively.

 On the S side, we diagonalize Eq.(\ref{eq:11}) and
 obtain
 the Bogoliubov quasi-particle states. The general form of
 the quasi-particle states on the S side is denoted by $\Phi^s(k_x^s,k_y)
 e^{i(k_x^sx+k_yy)}$ where $k_x^s$ is the longitudinal component of the wave vector on the S
 side. The 4-component spinor
 $\Phi^s(k_x^s,k_y)$ is called electron-like (hole-like) if the summation of
 the square of absolute values of the first two components is larger (lesser) than that
 of the last two components.  For each $\varepsilon$ and $k_y$, we can obtain four quasi-particle states.
 Two quasi-particle states  are picked out among
 four. The chosen states satisfy one of the following three
 conditions:
 (1) $k_x^s$ is real and positive and $\Phi^s(k_x^s,k_y)$ is
 hole-like;  (2) $k_x^s$ is real and negative and $\Phi^s(k_x^s,k_y)$ is
electron-like; (3) $k_x^s$ is complex and the imaginary part is
 negative. The last case corresponds to evanescent states near the
 interface.  By matching the wave functions of both sides at the interface x=0,
 we can obtain the reflection coefficients $r$ and $r_A$ for states $\Psi^e_{r}$ and
 $\Psi^h_{r}$, respectively. The quantum conductance through the
 S/N junction can be calculated using the Blonder-Tinkham-Klapwijk
 formula\cite{formula},
 \begin{equation}
 G=G_0\int_0^{\pi/2}(1-|r(eV,\alpha)|^2+n_h|r_A(eV,\alpha)|^2)\cos\alpha d\alpha
 \end{equation}
where $\alpha=tg^{-1}(\frac{k_y}{k_x})$ is the incident angle and
$G_0=\frac{4e^2}{h}N(eV)$ is the ballistic conductance of the
graphene sheet with density of states
$N(eV)=\frac{(E_F+eV)W}{v\pi}$ (W is the width of the graphene
sheet).
%EC
$n_h$ equals 1 if the hole state on the  N side is propagating,
and it is 0 if the state is evanescent.

For ease of comparing our results
 with the s-wave results in \cite{Beenakker1}, we depict
the normalized quantum conductance $G/G_0$ (as a function of bias
voltage) of the S/N junction with the S side being heavily doped
%EC graphene superconductor
superconducting graphene
for two cases, i.e., for $E_F>E_{\rm gap}^{sat}$ and
$E_F<E_{\rm gap}^{sat}$ in Fig.~\ref{fig:2}(a) and Fig.~\ref{fig:2}(b)
respectively, where $E_{\rm gap}^{sat}$ is the saturated gap for
$\mu\gg\Delta$ shown in Fig.\ref{fig:1}.

For $E_F>E_{\rm gap}^{sat}$, G  monotonically decreases in the
region $eV<\Delta$ and saturates to a constant value quickly as
$eV>\Delta$. The saturation value slightly decreases with $E_f$.
For $E_f<E_{\rm gap}^{sat}$, the line shape is similar to the s-wave
case, except that the unbiased conductance is nearly 2 instead of
$\frac{4}{3}$. It is noteworthy that G is always zero at the point
$E_f=eV$ in Fig.\ref{fig:2}b, since there is no Andreev hole
reflected back at this point for any angle of incidence.

The most remarkable difference between the $G/G_0-eV$ curves for the
conventional s-wave case\cite{Beenakker1} and the $d+id'$ wave case in
this paper is the value of the unbiased conductance, i.e., $\frac{4}{3}$
for s-wave and nearly 2 for our case. In ref\cite{Beenakker1}, the
lines are calculated in the large $U_0$ limit. To make things more
clear, we calculated the unbiased $G/G_0$ as a function of $U_0$
with several different choices of $E_f$ and $\Delta$ values. In
Fig.~\ref{fig:2}(c), we plot a typical comparison for three kinds of
pairing order parameters.
The results for the conventional
s-wave(\cite{Beenakker1}) and extended s-wave(\cite{p+ip}) cases show
little difference. For both cases the unbiased $G/G_0$ quickly
converges to the value of 4/3.  But for the $d+id'$ wave case considered here, $G/G_0$
decreases slowly from 2 and converges to $4/3$ after $U_0>10t$, which
is far beyond the single band edge. The most
%EC underlying
fundamental difference between the $d+id'$ pairing ansatz and
others is that it breaks time reversal symmetry and has an
emergent mixed s-wave and p-wave pairing at low energy. It would
be interesting to mention that in a recent conductance measurement
on a S/N/S structure, which is a realization of Andreev billiards,
the Andreev conductance is always peaked at zero voltage
bias\cite{expevidence}. This result is consistent with our
calculation and may shed new light on the superconducting pairing
symmetry of the graphene.

{\it Realization of $d+id'$-wave superconducting state in graphene:}
 It has been shown that the $d+id'$-wave superconducting state in graphene is a natural mean field solution
 in the presence of strong electron correlation\cite{d1}. Actually, the honeycomb
 lattice is closely related to triangular lattice(with same lattice rotational symmetry)
  in which the superconducting state is believed to be of
  $d\pm i d'$\cite{tri1,tri2}. Although the electron correlation in graphene is probably not strong enough to produce
a $d+id'$ superconducting by itself, it is possible to realize the
$d+id'$ superconducting state  by including the proximity effect
through a connection to another superconductor. For instance, let
us consider the geometry where the zigzag edge of graphene sheet
is laterally connected to the [110] direction of a d-wave pairing
superconductor on a square lattice (more specifically, a high Tc
cuprate superconductor). In the coordinate system we used to
discuss the symmetry of graphene, the order parameter in the $d$SC
side is $d_{xy}$. The proximity effect is therefore expected to
induce a $d_{xy}$ component in the graphene side near the
interface, which should be continually evolved into a $d+id'$
pairing symmetry  in the bulk due to the presence of
electron correlation.  Another possible realization is to put a
$d$SC on top of a graphene sheet. In the  long wave
length description, the effect of lattice mismatch is
irrelevant and
 $d+id'$ -wave superconducting state can also be induced. More radically, it is possible that even a
s-wave superconductor  may induce $d+id'$ order as well, as the
conductance measurement\cite{expevidence} that we mentioned
earlier indicates. The immediate consequence of the presence of
the $d+id'$ order is the spontaneous supercurrent along the
interface. A self-consistent study of such proximity effect will
be presented elsewhere.

In summary, we have shown  that a combination of s-wave and
$p+ip$-wave pairing order parameters  emerges at low energy in the
$d_{x^2-y^2}+id'_{xy}$ spin singlet pairing superconducting state
in graphene, which can be generated by electronic correlation as
well as induced through a proximity effect with a superconductor.
This mixture  of s-wave and $p+ip$  results in
%EC distinguished
distinctive
superconducting gap functions and novel behavior of the Andreev
conductance.

%\begin{figure}[tbh]
%\includegraphics[width=5cm,height=4cm]{linearcond-U.eps}
% Here is how to import EPS art
%\caption{The normalized quantum conductance of a S/N graphene
%junction for which the $x<0$ side is a d+id doped graphene
%superconductor and $x>0$ side is the normal graphene metal.
%$G_0=\frac{4e^2}{h}N(eV)$ is the ballistic conductance of the
%graphene sheet with density of states
%$N(eV)=\frac{(E_F+eV)W}{v\pi}$ where W is the width of the
%graphene sheet.}
% \label{fig:2}
%\end{figure}

\begin{acknowledgments}
Y. J. Jiang and J. P. Hu are supported by the National Science
Foundation (Grant No. PHY-0603759) and Natural Science Fundation of
Zhejiang province (Grant No.Y605167)of  China. D.-X. Yao is
supported by Purdue University. EWC is a Cottrell Scholar of
Research Corporation. HDC is supported by the DOE Award No.
DEFG02-91ER45439, through the Frederick Seitz Materials Research
Laboratory at UIUC.
\end{acknowledgments}

\end{document}